\documentclass{ws-ijqi}
\newcommand{\comment}[1]{}
\newcommand{\ket}[1]{\left|#1\right>}

\def\eps{\varepsilon}
\def\down {\downarrow}
\def\up {\uparrow}

\begin{document}

\markboth{Jos\'{e} Fernandez ,Tal Mor and Yossi Weinstein}
{Paramagnetic Materials and Practical Algorithmic Cooling for NMR-QC}

\catchline{}{}{}{}{}

\title{Paramagnetic Materials and Practical Algorithmic Cooling for 
NMR Quantum Computing\footnote{A conference version of this paper appeared in 
SPIE, volume 5105, pages 185-194 (2003)}}

\author{JOS\'{E} M.~FERNANDEZ}
\address{D\'{e}partment de g\'{e}nie informatique, \'{E}cole Polytechnique\\ Montr\'{e}al (Queb\'{e}c) Canada\\jose.fernandez@polymtl.ca}

\author{TAL MOR  and YOSSI WEINSTEIN}
\address{Faculty of Computer Science\\ Technion - Israel Institute of Technology\\ Haifa, Israel\\talmo@cs.technion.ac.il and yossiv@cs.technion.ac.il}


\maketitle

\begin{history}
\received{(1 July 2004)}
\end{history}

\begin{abstract}
Algorithmic Cooling is a method that uses novel data compression techniques
and simple quantum computing devices to improve NMR spectroscopy,
and to offer scalable NMR quantum computers. 
The algorithm recursively employs two steps. 
A reversible entropy compression of the \emph{computation quantum-bits (qubits)} 
of the system and an irreversible 
heat transfer from the system to the environment 
through a set of \emph{reset qubits} that reach thermal relaxation rapidly. 

Is it possible to experimentally demonstrate 
algorithmic cooling using existing
technology?
To allow experimental algorithmic cooling, the thermalization time
of the reset qubits must be much shorter than the thermalization time of
the computation qubits. However such thermalization-times ratios have
yet to be reported. 

We investigate here the effect of a paramagnetic salt on the 
thermalization-times ratio of computation
qubits (carbons) and a reset qubit (hydrogen). We show that the 
thermalization-times ratio is improved by approximately three-fold. 
Based on this result, an experimental demonstration of algorithmic 
cooling by thermalization and magnetic ions is currently performed
by our group and collaborators. 
\end{abstract}



\section{Introduction \label{introduction}}

NMR Quantum Computing (NMRQC) was proposed in 1997 by Cory,
Fahmy and Havel\cite{Fahmy}, and by Chuang and Gershenfeld\cite{Gershenfeld}.
They succeeded in implementing quantum algorithms on four-state quantum
systems (two spins). They used a liquid, an ensemble of identical
molecules. For encoding a quantum bit (qubit), the nuclear spin of one of the atoms
in the molecule is used. Each molecule is an independent quantum computer.
When measuring a qubit we get an average of the values of that qubit
over the ensemble. This technique is called ensemble quantum computing, and
in order to achieve quantum computation using this method, modifications
were introduced into the algorithms since they were developed for a single
quantum computer\cite{Gershenfeld,Roychowdhury}. To this day liquid
NMR is the most successful way of doing quantum computation. NMR
spectroscopists have succeeded in performing computation with up to
seven qubits in liquid NMR\cite{Factoring-15,7-qubits}, while other
successful methods have barely reached two-four qubits, and many
promising solid-state NMR methods are still struggling to encode a
single qubit. 

The probability of a spin-half which is coupled to a thermal bath and 
a constant magnetic field to be in the states $\ket{\up}$ and $\ket{\down}$ 
are given by
$P_{\up }=(1+\eps )/2 ,\, \, P_{\down }=(1-\eps )/2$.
The population bias $\eps$ is defined as
$\eps \triangleq \tanh [(\Delta E)/(2KT)]$,  
where $\Delta E$ is the energy gap between the two spin states, 
$K$ is the Boltzman coefficient and $T$ is the temperature in Kelvin.

Let each spin \( \frac{1}{2} \) particle be considered as a qubit.
 In a string of qubits (e.g. a molecule composed of several atoms with nuclear spin \( \frac{1}{2} \) ) the thermal state
density matrix of the i\( ^{th} \) qubit is
\begin{equation}
\label{Eq. single spin PPS}
\rho _{\eps_{0} ^{i}} =  \frac{1}{2}
\left( \begin{array}{cc}
1+\eps ^{i} & 0\\
0 & 1-\eps ^{i}
\end{array}\right).
\end{equation}
At room temperature the largest \( \eps  \) reached so far
is \( \eps \sim 10^{-5}. \) \( \eps  \) is also known as
the polarization bias, since it tells us the tendency of the system
to prefer one state over the other. The signal received from the sample
originates from this bias; the larger the bias, the stronger
the signal.
The state of the molecule is described by a tensor product of the single-qubit 
states, $\rho = \rho_{\eps_0^1} \otimes \cdots \otimes \rho_{\eps_0^{n}}$ 
for $n$ qubits. 

Unfortunately, the NMR quantum computing technique cannot work directly
with such mixed states. The algorithms are modified to use a pseudo-pure state
and then suffer from inherently bad 
signal-to-noise ratio (SNR). The SNR then decreases 
exponentially as the number of qubits 
grows\cite{Warren,DiVincenzo,SV,Cooling-I,Cooling-II}. 

\section{Algorithmic Cooling of Spins}

The scaling problem of NMR quantum computers can be resolved by
pre-cooling the qubits, which increases their bias, via two novel 
techniques recently developed. 
First, an adiabatic (reversible) 
cooling scheme has been introduced\cite{SV}, 
which solves the scaling problem using data compression tools
(that can actually be viewed as polarization compression tools).
This scheme does not suffer from the SNR problem anymore,
but is bounded by Shannon's bound on entropy-preserving compression, 
and therefore is limited in its practicality. 
Later on, {\em Algorithmic Cooling}\cite{Cooling-I}, a novel polarization 
compression technique, was suggested. It 
combines reversible compression together with thermalization steps;
by opening the system to the environment, and using thermalization (!!)
as a cooling mechanism, Shannon's bound is bypassed. 
An efficient and experimentally feasible algorithmic cooling
was then suggested\cite{Cooling-II}, with potentially important applications already in the 
near future:
improving SNR in NMR spectroscopy due to effective cooling of spins.
As far as we know, that work\cite{Cooling-II} provided the first near-future
application of quantum computing devices.

One can improve the SNR by cooling the system, increasing the magnetic field,
or increasing the sample size or number of sampling steps.
Another solution (that has some advantages over the above\cite{Cooling-II}
and can be combined with these other strategies as well) is 
an effective cooling of spins. This is a way to increase the spin bias without
cooling the system nor increasing the magnetic field.
Such a ``spin-cooling'' technique means 
taking the spin out of thermal equilibrium, then using the
improved polarization for spectroscopy 
before the spin goes back to its thermal equilibrium state. 
A well known example of such a spin-cooling 
technique is the polarization transfer between two
spins that have different polarization biases (e.g., proton and carbon
spins). Doing this, the proton is heated four times while the carbon is cooled accordingly.

The reversible 
(in-place) polarization compression\cite{SV} mentioned before is another
spin-cooling technique. It compresses 
the entropy from a few nuclear spins on a molecule 
to other spins on the same molecule.
Similar to the polarization transfer, this cooling method is also reversible, and therefore
it preserves the entropy contained in the system. 
It is limited due to the Shannon's bound: 
in order to have 
\( n_{j} \) pure-state qubits we would need to start 
with \(n_{0} = \ln(4) n_{j} \eps_{0} ^{-2}\) qubits\cite{SV}. 

In contrast, algorithmic cooling uses ``irreversible'' steps
as well.
In its simplest form,
each qubit is assigned a neighboring {\em reset qubit} 
to which the entropy is compressed. The reset qubits thermalize rapidly by 
radiating the compressed entropy out to the environment,
therefore losing heat and letting the entire spin system get cooler. 
Using a recursive algorithm based on these three 
steps, adiabatic compression, polarization transfer,
and thermalization, Shannon's bound
can be bypassed\cite{Cooling-I,Cooling-II} (by far), 
and low spin temperatures can be reached.

\section{Increasing Thermalization Times Ratio}

Polarization transfer is common in experimental NMR spectroscopy. 
Furthermore, an experimental demonstration of a reversible polarization compression
has already been performed\cite{BCS-exp} on molecules of C$_2$F$_3$Br.
Thus, the main 
limitation, preventing the demonstration of algorithmic cooling 
and the bypassing of Shannon's bound, 
is that the thermalization-times ratio between the 
computer qubits and the reset qubits should be large enough
so that the cooled computer qubits will not reheat while the reset 
qubits radiate the compressed entropy. We present and demonstrate now a novel
experimental technique
to improve the thermalization-times ratio in a real molecule,
opening the door for the first algorithmic cooling experiment. 

NMR pulse sequences are of typical time duration of milliseconds,
while the typical spin thermalization times in liquids are seconds.
Hence most of the sequence is spent on waiting for the reset qubits to thermalize. 
When a physical system is taken out of equilibrium, the relaxation
process back to equilibrium is always active. So not only do the warm
reset qubits cool down, but also the cooled spins warm back up. This
warming process is undesired, and therefore we need to find a way to
minimize it. If we succeed in increasing the \( T_{1} \) ratio between the
computational qubits and the reset qubits, the cooled qubits thermalization
would play a smaller rule during the reset qubits thermalization.

The use of paramagnetic salts for reducing the thermalization times
is a common practice in NMR spectroscopy.
Can it also be used to improve the thermalization-times ratio?
If we use protons as reset qubits and carbons as computation qubits, it may well be
that a paramagnetic salt will increase 
the \( T_{1} \) ratio.
A good reason to believe so is that,
in a typical carbon-chain molecule, the protons are more exposed
than the carbons to the environment solution. 

We took trichloroethylene  (TCE) 
\comment{(see Fig.~\ref{Fig: TCE}) 
\begin{figure}[th]
\centerline{\psfig{file=Trichloroethylene.eps,width=5cm}} 
\vspace*{8pt}
\caption{Tri-Chloro-ethylene labeled with two $^{13}$C. 
TCE has three nuclei with spin 1/2, the carbons and the proton.}
\label{Fig: TCE}
\end{figure}
}
with two \( ^{13}\textrm{C} \) nuclei (spin
1/2) used as computation qubits, 
and one proton (spin 1/2 as well) used as a reset qubit. The TCE was
dissolved in deuterated chloroform. 
We added the salt chromium(III)acetylacetonate
(AKA chromium 2,4-pentanedionate) to the solution at a concentration
of 233.2 mg/liter. 
Due to the physical structure of the TCE molecule, the proton has a
significantly stronger contact with the magnetic ions than the carbons.
The strong contact with the ions indeed decreases the thermalization
time, \( T_{1} \), of the proton significantly compared to the carbons,
achieving the goal of increasing the ratio of the \( T_{1} \)'s.
This is a very strong effect which can be observed by adding as little salt 
as we did. Naturally, the dephasing time, \( T_{2} \),
of all spins, also decreases. $T_2$ sets a boundary on the time available for 
computing. This is since any quantum information, which was obtained in the molecule, is lost due to dephasing.  Therefore the ionic salt concentration should 
allow enough time to perform a computation.
%

We used a BRUKER DMX-600MHz spectrometer at the University of Montreal.
The thermalization time (see Table~\ref{Tab:Magnetic-ions-effect})
\begin{table}[h]

\tbl{Thermal relaxation times in TCE before and after adding chromium salt to the solvent.\label{Tab:Magnetic-ions-effect}}
{\begin{tabular}{@{}llll@{}}\toprule
Label&                                           unsalted&    salted\\\colrule
\( T_{1}\left( C2\right)  \)&                    30.85 sec&     28.3 sec\\
\( T_{1}\left( C1\right)  \)&                    27.45 sec&     16.0 sec\\
\( T_{1}\left( H\right)   \)&                    5.460 sec&     1.88 sec\\
\(T_{1}\left(C2\right)/T_{1}\left(H\right)\)&    5.65&          15.05\\
\(T_{1}\left(C1\right)/T_{1}\left(H\right)\)&    5.03&          8.51\\\botrule
\end{tabular}}
\end{table}
of the proton decreased by 65.6\% while the nearest neighbor of the proton, C1,
and the next to nearest neighbor, C2, changed their thermalization times by 41.7\% and 8.3\% respectively. 
Meaning
that the ratio between the carbons and proton thermalization times 
went up by 69.2\% for C1 and by 166.4\% for C2. These ratios enable an experiment\cite{IMPOTENT} that can bypass Shannon's bound via some form of algorithmic cooling, in which two thermalization steps are performed.  

\section{Conclusions}

We showed here that paramagnetic materials can be used to increase the ratio of
 thermalization times of two spins on a molecule. This effect is most useful 
for algorithmic cooling, a novel process that improves signal-to-noise ratio in
 NMR spectroscopy, offering a solution to the scalability problem of NMR 
quantum computers.
We greatly improved the ratio of the thermalization times of
carbons and hydrogen, in trichloroethylene molecules, 
enabling their use for algorithmic cooling.
Based on the results presented here, the first
experiment presenting some form of 
algorithmic cooling is currently being performed\cite{IMPOTENT}.

Algorithmic Cooling has a patent pending No. 60/389,208.

\section*{Acknowledgments}
This work was supported in parts by the Israeli MOD and the Institute for Future Defense Research. We thank Yuval Elias and Matty Katz for helpful discussions.

\end{document}